# Trust in Human-AI Interaction: Scoping Out Models, Measures, and Methods

Takane Ueno
School of Engineering, Tokyo Institute of Technology Japan, ueno.t.ao@m.titech.ac.jp

Yuto Sawa
School of Engineering, Tokyo Institute of Technology Japan, sawa.y.ac@m.titech.ac.jp

Yeongdae Kim
School of Engineering, Tokyo Institute of Technology Japan, kim.y.ah@m.titech.ac.jp

Jacqueline Urakami
School of Engineering, Tokyo Institute of Technology Japan, urakami.j.aa@m.titech.ac.jp

Hiroki Oura
Department of Mathematics and Science Education, Tokyo University of Science Japan, houra@rs.tus.ac.jp

Katie Seaborn
School of Engineering, Tokyo Institute of Technology Japan, seaborn.k.aa@m.titech.ac.jp

## Abstract

Trust has emerged as a key factor in people's interactions with AI-infused systems. Yet, little is known about what models of trust have been used and for what systems: robots, virtual characters, smart vehicles, decision aids, or others. Moreover, there is yet no known standard approach to measuring trust in AI. This scoping review maps out the state of affairs on trust in human-AI interaction (HAII) from the perspectives of models, measures, and methods. Findings suggest that trust is an important and multi-faceted topic of study within HAII contexts. However, most work is under-theorized and under-reported, generally not using established trust models and missing details about methods, especially Wizard of Oz. We offer several targets for systematic review work as well as a research agenda for combining the strengths and addressing the weaknesses of the current literature.

## 1. Introduction

AI has been defined in the context of "the science and engineering of making intelligent machines, especially intelligent computer programs" [34:2]. People have different understandings about what "intelligence" is, for people or machines [53]. It may be described as futuristic and human-like, or it may be described as rational, like present-day AI [25]. AI-infused systems are becoming ubiquitous, increasingly used and recognized by non-experts. Common examples include Apple's virtual assistant and Internet search engine, Siri, a voice response application for smartphones and other devices that use the iOS operating system, and Softbank's humanoid robot Pepper, recently retired but established as a storefront figure and research tool. These modern AI-based technologies are built upon a foundation of smart systems designed to support human endeavors, notably automation in control systems [21,27]. Moreover, they are increasingly social in their embodiment [14,17,47]. Siri is conversational, albeit limited in scope, and Pepper strives to make eye contact and read the emotions of the people with whom it interacts. The same may not be said of their predecessors and intelligent cousins, planes and vehicles. However, in all cases, people are interacting with a machine of some perceived intelligence. We can also recognize an enduring pattern linked to this simulated intelligence: the intertwined notions of offloading human labor, especially cognitive processes, and human reliance on the systems that provide this labor in place of humans remain. As these AI-infused systems continue to be taken up by the public as well as specialists, there is great need to evaluate these underlying human factors that mediate their use.

Trust has been recognized as an important factor in interpersonal relationships between people, and potentially between people and machines. Researchers from a wide range of disciplines have examined the role of trust in mediating relationships between individuals, between individuals and organizations, and even between organizations [27]. But the importance of trust is not limited to interactions between people; it has also been highlighted as a key factor between people and computers [32,46,48], people and robots [9,19,39,45], and people and automation [21,27,36]. Recent work on people's impressions of interacting with AI-based technologies suggests that trust needs more attention. For instance, one attitudinal survey found that many people have a negative impression of AI-based technologies, with 48% Americans reporting that they would never get in a self-driving vehicle [50]. A Japanese study showed that, on average, more than 57% of people are uncomfortable working with AI in the workplace [35]. This hesitancy towards AI may have a variety of sources. One may be the unique nature of AI. Machine learning (ML), which is at the heart of current AI technologies, consists of algorithms and data training: a black-boxed and opaque state of affairs that has led to calls for Explainable AI or XAI [5]. Moreover, ML can involve relearning with new data and does not always produce the same output from the same input. The AI as a "mind" can be as unknowable and unpredictable as the minds of people, making interactions with AI-infused systems a matter of trust. Indeed, trust has been an important theme at AI-related conferences since the early days, and especially recently it has become a keyword in research on human-AI interaction (HAII) [3].

The high interest of researchers has led to an emerging and diverse body of work. However, little is known about what definitions, theories, and models of trust have been used and for what AI-infused systems. Recent literature reviews have started to trace out this body of work [16,52]. Glikson and Woolley [16] reviewed empirical studies on trust in all forms of AI, examining the factors that make up human trust in AI, but not how to assess these factors. Vereschak et al. [52] focused on the context of AI decision support, reviewing methods and providing practical guidelines for studying trust between humans and AI. What is missing is a general perspective that includes models and measures regardless of system type and context. Furthermore, we do not yet know of any standard approach to measuring trust when people interact with these AI-infused systems. As yet, it is not clear how to define trust or measure it in the context of HAII.

What is urgently needed is a map of the HAII research conducted so far to clarify what is meant by "trust," discover what models have been used, if any, and determine how trust has been evaluated. To this end, we have conducted a scoping review on trust in the HAII literature, from the earliest instance to the present day, covering a broad array of AI-infused systems. We asked: How is trust defined in HAII research? (RQ1) How is trust measured in HAII research? (RQ2) The main contributions of this work are threefold. First, we offer a comprehensive summary of the state of the art on how trust has been conceptualized and measured within HAII contexts. For this, we have provided a rigorous analysis of definitions, models, and measures. In particular, we identify two important issues that have not been addressed in previous reviews: empirical studies have generally not used established trust models; and there appears to be underreporting of a common approach to evaluation that has implications for working AI systems—Wizard of Oz. Second, we identify a list of targets for systematic review work. Finally, we suggest several objectives for empirical work. We have aimed to create a foundation for future work involving systematic reviews and empirical research.

## 2. METHODS

We conducted a scoping review in August 2021.Scoping reviews are used to determine the coverage of a body of literature on a particular topic [37]. They are meant to help researchers ask the right questions by quickly identifying and mapping the available evidence in that research area [2,4]. We used the guidelines provided by Munn et al. [37] and the extension [51] of the PRISMA checklist [31], the PRISMA-ScR (Preferred Reporting Items for Systematic reviews and Meta-Analyses extension for Scoping Reviews) [31,51], with modifications[1]. Our PRISMA flow diagram detailing the included and excluded papers at each stage of the review process is in Appendix A[2]. Our protocol was registered before data collection on July 19th, 2021[3].

### 1.1 Definitions

**AI-infused systems.** These are "systems that have features harnessing AI capabilities that are directly exposed to the end user" [1:1]. The AI capabilities are to make predictions, recommendations, or decisions influencing real or virtual environments, through learning, reasoning and self-correcting [15,57]. AI-infused systems is a broad category that can be applied to a range of technologies, including robots of all kinds, virtual agents, voice-based agents or agents with bodies, algorithms that provide even a small amount of interaction with people through an interface, and so on. We recognize that the degree to which people are aware that they are interacting with AI may vary. So, when we think of trust in HAII, we might be not only considering trust in the AI itself, but also trust in the entire AI-infused system. The factors include the representation, sociability, reputation and so on [49]. It also accounts for a common approach to deploying AI-based technologies in research, especially in HRI: Wizard of Oz (WoZ), where one or more people operate the system under study without the participant being aware of it [43]. Finally, it also accounts for scenarios wherein recordings are presented as a stimulus or vignettes are proposed as a thought experiment or anchor to jog memories. All these cases, as well as a range of AI-infused systems, are represented in the surveyed research.

---

[1] Note that we have adapted this protocol to account for the non-medical nature of the research surveyed. For instance, we did not use a structured abstract or the PICO model.

[2] https://osf.io/wrxq2

[3] https://osf.io/3hcr9

**Trust**. In a widely accepted definition, Mayer, Davis, and Schoorman [33:712] defined trust as "the willingness of a party to be vulnerable to the actions of another party based on the expectation that the other will perform a particular action important to the trustor, irrespective of the ability to monitor or control that other party." This definition includes "vulnerability" as a key element: we put ourselves at risk by delegating the responsibility for or consequences of an action to the trustee. This also applies to AI and automation [27]. Whether or not we can accept an agent—a person, a pet, a system's capabilities, an AI—depends on our trust in it. Whatever the form of intelligence, we can "disuse" it if our trust falls short of our perceptions of its capabilities, while we can also "misuse" it if our trust exceeds them. Put another way, an ideal state occurs when our trust in a given entity matches its capabilities.

## 1.2 Eligibility Criteria

We included full papers published in academic venues that reported on original human subjects research, including user studies, experiments, field studies, and so on. Papers also needed to feature results related to measuring trust in HAII in the abstract, title, and/or metadata. Papers that did not consider trust as a key variable were excluded. Papers were also excluded if they were not in English, Japanese, German, or Korean (the languages known within the team).

## 1.3 Information Sources and Search Strategy

We created a generic query structure that combined four dimensions: subject (types of AI-infused systems in HAII), domains of study, study focus, and review focus; see Table 1 for the full listing. The query was used to search in five major databases (ACM DL, IEEE Xplore, Scopus, Web of Science, and PsychInfo) in August 2021. A total of 330 papers were collected, excluding duplicates. Screening of the 330 papers was done in two stages. The first stage involved screening of only the titles and abstracts: all papers were screened by two authors, who identified 70 relevant papers. Additional literature was manually added, for a total of 102 new references. These were subjected to the same screening process as before by one researcher, and 23 relevant papers were identified. A total of 93 papers were then divided among four authors for a more detailed full-text screening according to the eligibility criteria. This resulted in 63 papers that were selected for data extraction. During data extraction, three papers were excluded for specific reasons, resulting in a final total of 60 papers included in the review. Excluded papers can be found in Appendix C[4].

## 1.4 Data Charting Process and Data Items

Data extraction was divided among four researchers. Information about the study (type of technology, methodology, participant demographics), trust definitions and models used, and metadata about independent and dependent variables (measurement, qualitative/quantitative, subjective/objective, response format, and timing of measurement, and validation) were extracted into a shared Google Sheet. The first author double-checked the data extractions.

---

[4] https://osf.io/vmq38/

**Table 1. General Query Structure for All Searches**

| Dimension[a] | Query Component |
| --- | --- |
| Subject | ai OR a.i. OR artificial intelligence* OR artificial agent* OR intelligent system* OR intelligent agent* OR artificially intelligent OR algorithm* OR intelligent algorithm* OR automation OR auto* agent* OR auto* system* OR intelligent automation OR intelligent robot* OR robot* assistant* OR embodied agent* OR embodied intelligent agent* OR virtual agent* OR virtual assistant* OR virtual intelligence* OR intelligent assistant* OR voice assistant* OR smart speaker* OR chatbot* OR speech recognition OR speech user interface* OR conversation* user interface* OR conversation* agent* OR computer vision OR machine vision OR machine perception OR image recognition OR expert system* OR deep learning OR neural network OR reinforcement learning OR anomaly detection OR decision-making system* OR recommend* system* OR self-driving car* OR self-driving vehicle* OR automated vehicle* OR automated driv* OR auto* vehicle* OR cooperative driv* OR co-operative driv* |
| Domains | human factor* OR user experience* OR ux OR human-robot interaction* OR hri OR human-computer interaction OR hci OR human-agent interaction OR hai OR human-machine communication OR hmc OR human-AI interaction OR HAII |
| Study Focus | user behavior* OR user study OR user studies OR user research OR user experience research OR ux research |
| Review Focus | trust* OR entrust* OR confiden* OR reliance OR rely OR relies OR reliable OR faith |

[a] Note that these dimensions are implicitly connected by AND. Some terms included the * qualifier to account for pluralization and different grammatical forms. For IEEE Xplore, the query was split and searched exhaustively because we could use only eight of the * qualifier at a time. Index terms were also used to filter results (e.g., "artificial intelligence") in some databases.

## 1.5 Data Analysis and Synthesis of Results

Descriptive statistics were used to summarize the quantifiable data. Thematic analyses were conducted on the trust definition data from an inductive, mixed semantic and latent, and constructionist orientation [7,18]. Two raters were involved. The lead rater developed the initial themes and then each rater separately coded all the data. Inter-rater reliability was assessed via Cohen's kappa using 0.8+ as a benchmark for agreement [26]; themes that did not meet this criterion were discarded. In borderline cases, disagreements were resolved through discussion. Similar themes were grouped together. Content analyses [6,25] of the trust measures data were conducted by two raters to categorize the data. Disagreements were discussed until resolved. A third rater resolved disagreements when needed. All other data (i.e., types of technology, independent variables) was categorized by one rater and checked by at least one other rater.

## 3. RESULTS

We now present the results according to our research questions. A list of included papers can be found in Appendix B[5].

---

[5] https://osf.io/y2zfj/

## 1.6 Defining Trust

*3.1.1 Trust definitions.* Less than half (46.7%, 28) of papers explicitly stated a definition of trust. 25% (15) used Lee and See's [27:51] definition: "the attitude that an agent will help achieve an individual's goals in a situation characterized by uncertainty and vulnerability." Our thematic framework of definitions is in Table 2. This framework shows that despite the diversity of definitions for trust, common patterns can be found in terms of trust as a concept, its features, and its stipulations.

*3.1.2 Trust models.* Only 23.3% (14) of papers explicitly referred to a model of trust, whether a human model or one created for or extended to machines. Of those that did, almost all used a unique model. One model was used in three papers: Hoff and Bashir's three-layered trust model [21]. Three papers used models that included emotional and cognitive components, but with different sources: Lewis and Weigert's social model of trust [29], Madsen and Gregor's Model of Human-Computer Trust Components [32], Cook and Wall's organizational model [11], and Johnson and Grayson's model in service relationship contexts [24]. Other models included Muir's five-factor model [36] and Lee and Moray's Automation Trust Model [28].

## 1.7 Measuring Trust

Across 60 papers, there were 10,342 participants: 4210 men (40.7%), 3950 women (38.2%), and 17 unidentified or labelled as "other" (0.2%). This shows that men and women were roughly equal in number, while it is unclear how those who did not select these gender/sex categories identify. 13.3% (8) did not report on gender/sex. The mean age was 27.8 (MED=25.6, SD=7.9, IQR=32), indicating a generally younger adult age range. However, 28.3% (17) did not report on age or reported age in a format that could not be transformed for quantitative analysis. Country and race/ethnicity were reported in diverse ways; only 28.4% (34) reported on these demographics. The most common categories were US (16.7%, 10) and China (8.3%, 5); however, it is likely that these numbers are inaccurate due to diverse reporting or lack of reporting.

Our survey revealed three distinct orientations to measuring trust in HAII research: measuring trust directly with real AI-infused systems in interaction contexts (53%, 32); measuring trust directly through simulated experiences with AI-infused systems, such as through WoZ (27%, 16); and measuring trust indirectly through impressions or imagined scenarios, such as by evaluating attitudes or presenting vignettes (25%, 15). 8% (5) included multiple orientations within or across studies. There were four studies that seemed to use WoZ; only one explicitly stated that WoZ was used. Three used simulations, and one used video-based scenario.

Most studies (85%, 51) involved some type of independent variable manipulated by the researchers or a grouping variable. These could be grouped into seven categories: reliability (17.3%, 9), predictability (3.8%, 2), explainability (30.8%, 16), trust calibration (17.3%, 9), agent type (53.8%, 28), demographics (11.5%, 6), and difficulty (9.6%, 5).

**Table 2. Thematic Framework of Trust as Defined in HAII Research; Frequency Counts in Parentheses**

| Theme | Sub-theme | Definition | Example |
|---|---|---|---|
| Concept Characteristics | Multidimensional (10) | A multifaceted concept made up of dimensions that may be considered separately or together. | "Trust is a multi-dimensional concept … the dimensionality of trust ..." [22:3] |
| | Anthropic (22) | A human perception and human-centered. | "... a critical moderator of the relationship between humans and machines …" [8:2] |
| | Attitudinal (19) | An attitude, belief, or perspective, or frame of mind. | "... the attitude that an agent will help achieve an individual's goals …" [20:4] quoting [27:51] |
| | Cognitive (4) | A cognitive process of thinking and judging. | "... reasons and arguments ..." [40:2] |
| | Affective (4) | An emotional response or based in feelings. | "...the feeling of a person towards an agent …" [40:2] |
| | Context Dependent (3) | Mediated by the context of use. | "... strongly dependent on the situational context" [54:9] |
| | Longitudinal (6) | Emerges over time rather than momentary. | "… a long term relationship …" [42:543] |
| Features of Trust in Action | Benevolent (5) | Results in a positive outcome. | "... actions prejudicial to their well-being …" [12:509] |
| | Reliable (13) | Re/actions are predictable, steady, and transparent. | "... primarily based on the predictability of the system's behavior." [10:002884] |
| | Honest (3) | Re/actions are made with integrity and authenticity. | "... based on evaluations of the focal technology's integrity …" [22:3] |
| | Competent (8) | Re/actions demonstrate high performance. | "... perceived technical competence …" [30:7] |
| | Assistive (16) | Re/action serves to assist in achieving one's goal or meeting one's expectations. | "... another will follow through on your behalf …" [55:244] |
| Stipulations of Trust | Voluntary (7) | Trust is voluntary, based on willingness. | "...willingness to rely on [it] ..." [38:194] |
| | Vulnerability (16) | Trust means accepting the uncertainty that one may come into harm or experience loss. | "... a situation characterized by uncertainty and vulnerability …" [13:2] quoting [27:51] |

AI-based technologies were classified into general categories. These included: systems related to automated driving (31%, 19), recommender systems such as product recommendation in e-commerce (15%, 9), decision support systems including computer decision aid and robot decision aid (13%, 8), and robots that act autonomously (13%, 8). Others included medical AI, chatbots, and facial recognition systems (28%, 17). A variety of contexts and scenarios were used, such that it was difficult to develop general categories for reporting.

In total, 95 trust measures across three major categories—General Trust, Distrust, and Specific Trust—were identified; see the full table in Appendix D[6]. General Trust refers to overall trust, involving multiple concepts combined in a single measure. Distrust refers to a lack of trust. Specific Trust includes other measures that are recognized as part of trust in the literature (e.g., Acceptance). Almost all (89%, 85) measures were evaluated through questionnaires. 72% (68) measured General Trust. Of these, 91% (62) used questionnaires, with two qualitative interviews and four observations reported. 22% (21) measured Specific Trust, with most (90%, 19) using questionnaires and the other two measured by observation. Distrust was reported to be measured in 6% (6 cases); of these, 4 were by questionnaire and 2 were by observation. 44% (42) were original measurements made by the researchers and the remaining 56% (53) were existing or inspired by existing measurements. The most cited (13 studies) was the 12-item Trust in Automation Scale [28]. Two approaches to measuring distrust were found: one where the value for distrust was inverted and used as a single trust scale, and another where trust and distrust were reported as separate values. The remaining instruments were unique. In addition, 21% (18) of the questionnaires had only one item; in 4 cases the number of items was not reported. Trust was measured at different times. 11% (10) measured trust only before, 71% (67) only after, 6% (6) before and after, and 8% (8) during the study. Others (2%, 2) were reported to have been measured multiple times throughout the experiment over multiple days. 45% (43) of instruments were validated. 44% (42) assessed internal reliability.

## 4. DISCUSSION

Trust is a multidimensional factor being studied across a range of AI-based technologies. Yet, it is undertheorized and potentially limited in scope. Less than half of the surveyed papers reported on an explicit definition of trust, and one quarter of those relied on a single definition, by Lee and See [27]. At the same time, our thematic analysis revealed several patterns across the available definitions for trust. Trust has been characterized as anthropic, or human-centered rather than agent-neutral, attitudinal rather than behavioral, requiring vulnerability, being about receiving expected assistance towards one's goals, and reliable over time. Most papers did not use a specific model. A few referenced Hoff and Bashir's three-layer model of trust [21]. Others used a two-factor structure of emotional and cognitive components of trust [11,24,29,32]. Future work may seek to trace commonalities between these models and evaluate their generalizability to various types of AI-infused systems. Aside from these cases, there was little consensus on models used. This diversity may be seen as a weakness or a strength; the choice for researchers is great, but it is not clear if such a selection of choices is necessary. Indeed, the patterns shown through the thematic analysis results suggest that this is not the case. Nevertheless, specific models may be useful for specific dimensions of trust, which future work can explore.

The demographics and measurement results revealed inadequate reporting. Demographics, especially in terms of race, ethnicity, and nationality, may be particularly important for trust, given that previous research has found cross-cultural effects and a Western bias, e.g., [44,56]. Basic demographics should be reported in forms that can be compared. Research that includes non-Western models and measures of trust should be explored. Comparing the definitions of trust and measures reveal some unexpected patterns. Even though reliability was a key component of most definitions of trust, only about one-sixth of papers used it as a grouping or manipulation variable. At the same time, agent type was manipulated by over half of studies, yet technology-specific features were not represented in trust definitions and most models used. One grouping factor that may be specific to trust is trust calibration, generally described as a prime or cue that influences the participant's perception of trust in the technology. Future systematic review work can assess the degree that outcomes change depending on the kinds and levels of trust calibration cues studied.

---

[6] https://osf.io/uvnfj/

A wide variety trust measurements were found. Even so, several studies (13 of 60) used the Trust in Automation Scale by Jian et al. [23]. This instrument, in name and practice, has been widely used to measure trust in automation. However, given the potential differences between automation and nonautomation technologies, its applicability for AI-based technologies in general needs to be properly assessed. Furthermore, our findings show that more than half of the measures used have neither been validated nor assessed for internal reliability. In light of this, there is an urgent need to create a uniform and valid measure of trust for AI-based technologies.

A long-standing challenge is the use of WoZ. Part of the discourse on WoZ has focused on its *invisibility*, with authors not clearly stating whether they used a WoZ approach. Indeed, the organizers of the HRI conference have recently[7] strongly encouraged authors to do so, based on Riek's reporting guidelines [43]. As our results on orientations to measuring trust in HAII reveal, nearly half of papers purportedly used a working system, with about one-third reporting use of WoZ. Yet, we cannot be sure that all the former half did not involve WoZ. Aside from the general issues covered by Riek for HRI research (e.g., lack of Wizard training, Wizard errors, Wizards being "too real to be a machine"), use of WoZ is likely to play a role in trust outcomes. Trust, according to our survey results, is often defined as predictable behavior. But behavior can vary, even in general ways. Humans may be the most variable, but machine learning algorithms can be unpredictable, while other kinds of algorithms, such as those used by commercial robots that may not change over time, may be too predictable. Going forward, it would be useful to compare WoZ and actual technologies of various kinds, to determine how great or small the difference in trust outcomes may be on a case-to-case basis.

Scoping reviews often lead to systematic reviews [37,41]. Yet, we were unable to identify many common points for comparison. For instance, no specific technology was used in multiple studies, despite the popularity of such robotic platforms as Aldebaran-SoftBank's Nao. Moreover, most studies did not use a model of trust, making comparisons difficult and limiting empirical contributions to theory. Additionally, the plethora of contexts and measures limit generalizability. Future work will need to assess the same technologies, contexts, and scenarios using the same trust measures to set the stage for future systematic review work. One exception was the general types of technologies studied: automated driving assistants, recommender systems, and decision support systems—robotic ones. A systematic review may focus on the same measure for a given general type even where the specific technology differs.

---

[7] See for instance the request for authors submitting to HRI 2021: "if a Wizard-of-Oz paradigm was used, a detailed description of the robot, wizard, user, etc.": https://humanrobotinteraction.org/2021/full-papers

## 1.8 Limitations

We restricted our scoping review to records that used "trust" and other keywords in the abstract, title, and metadata. We recognize that not all viable papers did so. While this is fine for scoping review work, systematic work should consider the full text for query searches and also use a quality assessment tool [41].

## 5. CONCLUSION

Artificial intelligence, or its simulation, has advanced to the point of providing realistic and meaningful interactions with people. Trust has emerged as an important factor, grounded in automation studies and other forerunners to modern intelligent agents, tools, and systems. As this scoping review has shown, explorations of the nature of trust and its measurement within HAII contexts is a burgeoning but underdeveloped topic of research. We have offered points of departure for systematic review work, deeper theorizing, revaluations of methods and measures, and trajectories for empirical work. We trust that this scoping review may be used as a guide to steer the course.